\documentclass[dvips]{article}
\usepackage{graphicx}
\usepackage{amsbsy}
\newcommand{\doublespacing}{\let\CS=\@currsize\renewcommand{\baselinesstrech}
{2.0}\tiny\CS} \textwidth 16cm
\newcommand{\bd}{
\begin{document}}
\newcommand{\ed}{\end{document}}
\newcommand{\bc}{\begin{center}}
\newcommand{\ec}{\end{center}}
\newcommand{\bfr}{\begin{flushright}}
\newcommand{\efr}{\end{flushright}}
\newcommand{\lt}{\left}
\newcommand{\rt}{\right}
\newcommand{\vs}{\vspace}
\newcommand{\hs}{\hspace}
\newcommand{\beq}{\begin{equation}}
\newcommand{\eeq}{\end{equation}}
\newcommand{\lb}{\linebreak}
\newcommand{\pb}{\pagebreak}
\newcommand{\mb}{\makebox}
\newcommand{\fb}{\framebox}
\newcommand{\mc}{\multicolumn}
\newcommand{\ben}{\begin{enumerate}}
\newcommand{\een}{\end{enumerate}}
\newcommand{\bit}{\begin{itemize}}
\newcommand{\eit}{\end{itemize}}
\newcommand{\ol}{\overline}
\newcommand{\un}{\underline}
\newcommand{\lefq}{\lefteqn}
\newcommand{\ba}{\begin{array}}
\newcommand{\ea}{\end{array}}
\newcommand{\beqa}{\begin{eqnarray}}
\newcommand{\eeqa}{\end{eqnarray}}
\newcommand{\beqas}{\begin{eqnarray*}}
\newcommand{\eeqas}{\end{eqnarray*}}
\newcommand{\bfg}{\begin{figure}}
\newcommand{\efg}{\end{figure}}
\newcommand{\bds}{\begin{displaymath}}
\newcommand{\eds}{\end{displaymath}}
\newcommand{\btb}{\begin{tabbing}}
\newcommand{\etb}{\end{tabbing}}
\newcommand{\para}{\parallel}
\newcommand{\pad}{\partial}
\newcommand{\nn}{\nonumber}
\newcommand{\la}{\leftarrow}
\newcommand{\ra}{\rightarrow}
\newcommand{\lgla}{\longleftarrow}
\newcommand{\lgra}{\longrightarrow}
\newcommand{\La}{\Leftarrow}\newcommand{\Ra}{\Rightarrow}
\newcommand{\Lra}{\Leftrightarrow}
\newcommand{\Lgla}{\Longleftarrow}
\newcommand{\Lgra}{\Longrightarrow}
\newcommand{\bm}{\boldmath}
\newcommand{\lan}{\langle}
\newcommand{\ran}{\rangle}
\renewcommand{\a}{\alpha}
\renewcommand{\b}{\beta}
\newcommand{\g}{\gamma}
\newcommand{\G}{\Gamma}
\renewcommand{\d}{\delta}
\newcommand{\eps}{\epsilon}
\newcommand{\Th}{\Theta}
\newcommand{\s}{\sigma}
\newcommand{\lam}{\lambda}
\newcommand{\D}{\Delta}
\newcommand{\vare}{\varepsilon}
\newcommand{\pr}{\prime}
\newcommand{\ro}{\rho}
\newcommand{\nab}{\nabla}
\newcommand{\m}{\mu}
\newcommand{\n}{\nu}
\newcommand{\Sg}{\Sigma}
\newcommand{\p}{\pi}
\newcommand{\R}{I\!\!R}
\newcommand{\om}{\omega}
\newcommand{\Om}{\Omega}
\newcommand{\ze}{\zeta}
\newcommand{\vart}{\vartheta}
\newcommand{\tri}{\triangle}
\newcommand{\f}{\frac}
\newcommand{\iny}{\infty}
\newcommand{\pro}{\propto}
\bd
\bc
 \Large{\textbf{Potential algebra approach to position dependent mass Schr\"odinger equations}} \ec
\vspace{.5cm}

\bc{\it T. K. Jana{\footnote {e-mail :
tapasisi@gmail.com}\\
Department of Mathematics\\
R.S.Mahavidyalaya, Ghatal 721212, India\\
 
\vspace{.2cm}

P. Roy{\footnote{e-mail : pinaki@isical.ac.in}}\\
Physics \& Applied Mathematics Unit \\
Indian Statistical Institute \\
Kolkata 700108, India.}} \ec \vs{4.5cm}

\bc {\large {\un{Abstract}}} \ec 
It is shown that for a class of position dependent mass Schr\"odinger equation the shape invariance condition is equivalent to a potential symmetry algebra. Explicit realization of such algebras have been obtained for some shape invariant potentials.\\\\
\vs{4.5cm}
\pb
\section{Introduction}
Position dependent mass Schr\"odinger equation (PDMSE) has attracted a lot of attention during the past few years. This is because of the possible applications of PDMSE in a variety of fields like describing the dynamics of electrons in many condensed matter systems, such as compositionally graded crystals \cite{Geller}, quantum dots \cite{Serra} and quantum liquids \cite{Arias}, in the determination of the electronic
properties of semiconductors \cite{Bastard}, $^{3}$ He cluster \cite{Barranco} etc. and also due to intrinsic interest in such systems.
Because of the importance of exact solutions, there have been a growing interest in obtaining the exact solutions of PDMSE in different contexts \cite{pdm}. Exact solutions of PDMSE's can be obtained using different methods \cite{Dekar1}. Among the various techniques used so far algebraic ones like the Lie algebraic approach \cite{roy,roy1,koc1,koc2,quesne1} is particularly interesting since it produces not only the solutions but also reveals the symmetry of the problem. On the other hand another algebraic technique, namely, the shape invariance approach has been found to be extremely useful in obtaining exact solutions of PDMSE \cite{samani,quesne,quesne2}. In this context a natural question is the following: Is the shape invariance method related to Lie algebraic ones?  
It has been shown recently that the answer to this question is in the affirmative in the case of constant mass Schr\"odinger equation \cite{fukui,balan,gan}. It may be noted that because of the presence of a non constant mass the shape invariance condition in the case of PDMSE is different from the constant mass case and here our objective is to examine whether or not potential symmetry algebras can be found for shape invariant potential within the context of PDMSE.  Here we shall show the existence of potential algebras for a couple of translationally shape invariant potentials while the results for the other cases will be presented in the form of table. 
 
\section{Shape invariant PDMSE}\label{sipdmse}
Let us begin with the deformed Schr\"odinger equation given by \cite{quesne1}
\beq
H\psi=[\pi^2+V(x)]\psi=E\psi \label{sch1}\eeq
where the deformed momentum $\pi$ is given by
\beq
\pi=-i \sqrt{f(x;\boldsymbol{\alpha})}\frac{d}{dx}\sqrt{f(x;\boldsymbol{\alpha})}\label{cr1} \eeq 
and $f(x;\boldsymbol{\alpha})$ is the deforming function which is assumed to be real and positive and $\boldsymbol{\alpha}$ is a set of parameters.\\
Eq.$(\ref{sch1})$ can be written in the form
\beq
\left(-\frac{d}{dx}\frac{1}{M(x;\boldsymbol{\alpha})}\frac{d}{dx}+V_{eff}(x;\boldsymbol{\alpha})\right)\psi =E\psi \label{sch2}\eeq
where $M(x;\boldsymbol{\alpha})=f^{-2}(x;\boldsymbol{\alpha})$. Thus Eq.$(\ref{sch2})$ can be identified as a PDMSE \cite{levy} with a mass function $M(x;\boldsymbol{\alpha})$ and an effective potential given by 
\beq
V_{eff}(x;\boldsymbol{\alpha})=V(x)-\frac{1}{2}f(x;\boldsymbol{\alpha}) f''(x;\boldsymbol{\alpha}) -\frac{1}{4}f'^2(x;\boldsymbol{\alpha})\label{veff}\eeq 
Now we consider two operators $A^-$ and $A^{+}$ of the form
\beq
A^-(\boldsymbol{\alpha};\boldsymbol{\lambda})=\sqrt{f(x;\boldsymbol{\alpha})}\frac{d}{dx}\sqrt{f(x;\boldsymbol{\alpha})}+W(x;\boldsymbol{\lambda}) ,~~A^{+}(\boldsymbol{\alpha};\boldsymbol{\lambda})=-\sqrt{f(x;\boldsymbol{\alpha})}\frac{d}{dx}\sqrt{f(x;\boldsymbol{\alpha})}+W(x;\boldsymbol{\lambda})\label{aadag}\eeq
where $\boldsymbol{\lambda}$ is a set of parameters.
Then by construction $H_{-}=A^{+}A^-$ and $H_{+}=A^-A^{+}$ are isospectral partners. The potentials appearing in $H_{\pm}$ are then given by
\beq
V_{\pm}=W^{2}(x;\boldsymbol{\lambda}) \pm f(x;\boldsymbol{\alpha})W^\prime(x;\boldsymbol{\lambda})\label{vpm1}\eeq

These two partner potentials are shape invariant if
\beq
W^{2}(x;\boldsymbol{\lambda_1}) + f(x;\boldsymbol{\alpha})W'(x;\boldsymbol{\lambda_1})=W^{2}(x;\boldsymbol{\lambda_2}) - f(x;\boldsymbol{\alpha})W'(x;\boldsymbol{\lambda_2})+R(\boldsymbol{\lambda_1})\label{sic1}\eeq
where $\boldsymbol{\lambda_{2}}$ is a function of $\boldsymbol{\lambda_{1}}$ and $R$ is a function independent of $x$. 
Using the operators defined in (\ref{aadag}), the shape invariance condition (\ref{sic1}) can be written  as
\beq
A^{-}(\boldsymbol{\lambda_1})A^+(\boldsymbol{\lambda_1})=A^+(\boldsymbol{\lambda_2})A^{-}(\boldsymbol{\lambda_2})+R(\boldsymbol{\lambda_1})\label{sic2}\eeq

\section{Algebraic structure of translational shape invariance} 
Here we shall be considering translationally shape invariant potentials i.e, $\boldsymbol{\lambda_{n}}=\boldsymbol{\lambda_{1}}+(n-1)\boldsymbol{\eta}$. 
To establish the algebraic structure of such systems we first consider the following operators \cite{fukui,balan}
\beq
\ba{l}
B_+(\boldsymbol{\lambda_{1}})=A^{\dagger}(\boldsymbol{\lambda_{1}})T(\boldsymbol{\lambda_{1}})\\
B_-(\boldsymbol{\lambda_{1}})=B_{+}^{\dagger}(\boldsymbol{\lambda_{1}})=T^{\dagger}(\boldsymbol{\lambda_{1}})A(\boldsymbol{\lambda_{1}})
\ea
\label{bbdag}\eeq
where the operator $T(\boldsymbol{\lambda_{1}})$ and its conjugate  $T^{\dagger}(\boldsymbol{\lambda_{1}})$ are given by
\beq
T(\boldsymbol{\lambda_{1}})=e^{\boldsymbol{\eta} \frac{\partial}{\partial \boldsymbol{\lambda_{1}}}}~~\mbox{and}~~T^{\dagger}(\boldsymbol{\lambda_{1}})= e^{-\boldsymbol{\eta} \frac{\partial}{\partial \boldsymbol{\lambda_{1}}}}\label{ttdag}\eeq

In terms of the operators $B_+(\boldsymbol{\lambda_{1}})$ and $B_-(\boldsymbol{\lambda_{1}})$ the Hamiltonian $H_-$ can be rewritten as
\beq
H_-=A^{\dagger}A=A^{\dagger}(\boldsymbol{\lambda_{1}})T(\boldsymbol{\lambda_{1}})T^{\dagger}(\boldsymbol{\lambda_{1}})A(\boldsymbol{\lambda_{1}})=B_+(\boldsymbol{\lambda_{1}})B_-(\boldsymbol{\lambda_{1}})\label{hm}\eeq
Now using (\ref{sic2}) and the identity $R(\boldsymbol{\lambda_n})=T(\boldsymbol{\lambda_{1}})R(\boldsymbol{\lambda_{n-1}})T^{\dagger}(\boldsymbol{\lambda_{1}})$ it can be shown that
\beq
\left[B_-,B_+\right]=R(\boldsymbol{\lambda_0})\label{bmbpc}\eeq
Also the operators $B_-,$ and $B_+$ satisfy the following relations
\beq
\ba{l}
R(\boldsymbol{\lambda_n})B_+ =B_+R(\boldsymbol{\lambda_{n-1}})\\
R(\boldsymbol{\lambda_n})B_- =B_-R(\boldsymbol{\lambda_{n+1}})\\
\left[H_-,B_{+}^{n} \right]=[R(\boldsymbol{\lambda_1})+R(\boldsymbol{\lambda_2})+.......+R(\boldsymbol{\lambda_n})]B_{+}^{n}\\
\left[H_-,B_{-}^{n} \right]=-B_{-}^{n}[R(\boldsymbol{\lambda_1})+R(\boldsymbol{\lambda_2})+.......+R(\boldsymbol{\lambda_n})]\\
\ea
\label{relations}\eeq
For unbroken supersymmetry the ground state of $H_-$ satisfies the condition
\beq\
B_{-}\left|\psi_0\right\rangle=0\label{gs}
\eeq 
Then from Eq.$(\ref{relations})$ and Eq.$(\ref{gs})$ it follows that
\beq
H_-(B_{+}^{n}\left|\psi_0\right\rangle)=[R(\boldsymbol{\lambda_1})+R(\boldsymbol{\lambda_2})+.......+R(\boldsymbol{\lambda_n})](B_{+}^{n}\left|\psi_0\right\rangle)\label{es}\eeq
Thus $B_{+}^{n}\left|\psi_0\right\rangle$ is an eigenstate of the Hamiltonian $H_-$ with the eigenvalue $R(\boldsymbol{\lambda_1})+R(\boldsymbol{\lambda_2})+.......+R(\boldsymbol{\lambda_n})$.\\

From the first two relations of Eq.$(\ref{relations})$ one gets the following commutation relations
\beq
\ba{l}
\left[B_{+},R(\boldsymbol{\lambda_0})\right]=[R(\boldsymbol{\lambda_1})-R(\boldsymbol{\lambda_0})]B_{+}\\
\left[B_{+},\left[B_{+},R(\boldsymbol{\lambda_0})\right]\right]=\left\{[R(\boldsymbol{\lambda_2})-R(\boldsymbol{\lambda_1})]-[R(\boldsymbol{\lambda_1})-R(\boldsymbol{\lambda_0})]\right\}B_{+}\\
\ea
\label{comr1}\eeq
and so on. However if the difference $R(\boldsymbol{\lambda_n})-R(\boldsymbol{\lambda_{n-1}})$ becomes a (non zero) constant then only the first commutator of (\ref{comr1}) survives.

\section{Examples}
In this section we shall consider a few examples to illustrate the method described in the previous section.\\ 

\textbf{Example 1.} Let us first take $W(x;\boldsymbol{\lambda})=\lambda_1 x$ and $f(x;\boldsymbol{\alpha})=1+\alpha x^2$, $(\a>0)$ in Eq.$(\ref{vpm1})$, so that
\beq
V_{\pm}=\lambda_1(\lambda_1 \pm \alpha)x^2\pm \lambda_1\label{vpm}
\eeq  
Using (\ref{sic1}) it can be shown that $V_-$ is a shape invariant potential with 
\beq
\lambda_n=\lambda_{n-1}+\alpha ~~ \mbox{and}~~R(\lambda_n)=2\lambda_n+\alpha \label{rn}
\eeq
Hence the energy spectrum is given by 
\beq
E_{n}^{-}=\sum_{i=1}^{n}R(\lambda_i)=2n \lambda_1+n^2 \alpha
\eeq 
Here 
\beq
\ba{l}
B_{+}(\lambda_1)=\displaystyle\left(-\sqrt{1+\alpha x^2}\frac{d}{dx}\sqrt{1+\alpha x^2}+\lambda_1 x\right)e^{\alpha \frac{\partial}{\partial \lambda_1}}\\
B_{-}(\lambda_1)=\displaystyle e^{-\alpha \frac{\partial}{\partial \lambda_1}}\left(\sqrt{1+\alpha x^2}\frac{d}{dx}\sqrt{1+\alpha x^2}+\lambda_1 x\right)\\
\ea
\label{bpbm1}\eeq

Now from (\ref{bmbpc}) we have
\beq
\left[B_{-}(\lambda_1),B_{+}(\lambda_1)\right]=R(\lambda_0)=2\lambda_0+\alpha=2\lambda_1-\alpha\label{bpm}
\eeq
Also from (\ref{rn}) it can be shown that
\beq
R(\lambda_n)-R(\lambda_{n-1})=2\alpha=constant
\eeq
so that
\beq
\left[B_{\pm},R(\lambda_0)\right]= \pm 2\alpha B_{\pm}\label{bpm1}
\eeq 
while the other commutators of Eq.$(\ref{comr1})$ vanish. Thus the commutation relations (\ref{bpm}) and (\ref{bpm1}) constitute a finite dimensional potential algebra for the potential (\ref{vpm}).\\\\ 

\textbf{Example 2.} As another example we consider a two parameter potential for which $W(x;\lambda,\mu)=\lambda_1 \tan x+\mu_1 \sec x$ and $f(x;\boldsymbol{\alpha})=1+\alpha \sin x$, $(-1<\a<1)$. Then from Eq.$(\ref{vpm1})$ we get
\beq
V_{\pm}=[\lambda_1(\lambda_1 \pm 1)+\mu_1(\mu_1 \pm \alpha)]\sec^2 x+[\mu_1(2\lambda_1 \pm1)\pm \alpha \lambda_1]\sec x \tan x-\lambda_{1}^{2} \mp \alpha \mu_1\label{ex2}
\eeq

Here also $V_-$ is a shape invariant potential with
\beq
\ba{l}
\lambda_n=\lambda_{n-1}+1,~~~~~\mu_n=\mu_{n-1}+ \alpha\\\\
R(\lambda_n,\mu_n)=2 \lambda_n+1-\alpha(2\mu_n+\alpha)\\
\ea
\eeq
Therefore the energy spectrum is given by 
\beq
E_{n}^{-}=\sum_{i=1}^{n}R(\lambda_i,\mu_i)=n^2(1-\alpha^2)+2n(\lambda_1-\alpha \mu_1)
\eeq
Since there are two independent parameters in the potential we take $T(\lam,\mu)=e^{\f{\partial}{\partial\lam_1}+\a\f{\partial}{\partial\mu_1}}$  and so the operators $B_{+}$ and $B_{-}$ are given by
\beq
\ba{l}
B_{+}(\lambda_1,\mu_1)=\displaystyle \left[-\sqrt{1+\alpha \sin x}\frac{d}{dx}\sqrt{1+\alpha \sin x}+(\lambda_1 \tan x+\mu_1 \sec x)\right]e^{\frac{\partial}{\partial \lambda_1}+\alpha \frac{\partial}{\partial \mu_1}}\\
B_{-}(\lambda_1,\mu_1)=\displaystyle e^{-\frac{\partial}{\partial \lambda_1}-\alpha \frac{\partial}{\partial \mu_1}}\left[\sqrt{1+\alpha \sin x}\frac{d}{dx}\sqrt{1+\alpha \sin x}+(\lambda_1 \tan x+\mu_1 \sec x)\right]\\
\ea
\label{bpbm2}\eeq
In this case we have 
\beq
\left[B_{-},B_{+}\right]=R(\lambda_0,\mu_0)=2(\lam_1-\a\mu_1)+\a^2-1\label{bpm3}
\eeq
Also $R(\lambda_n,\mu_n)-R(\lambda_{n-1},\mu_{n-1})=2(1-\alpha^2)$ and consequently
\beq
\left[B_{\pm},R(\lambda_0,\mu_0)\right]= \pm 2(1-\alpha^2) B_{\pm}\label{bpm4}
\eeq 
So the commutators (\ref{bpm3}) and (\ref{bpm4}) constitute the finite dimensional potential algebra for the potential $V_-(x;\lam,\mu)$ in (\ref{ex2}).\\ 

So far we have considered examples for which supersymmetry is unbroken. However the present method can also be extended to some systems with broken supersymmetry. To show this let us consider the first example of Table 1. In this case 
\beq
H_{\mp}=\pi^2+\f{\lam_1(\lam_1\pm1)}{x^2}+\mu_1(\mu_1\mp\a)x^2+2\lam_1\mu_1\pm\a\lam_1\mp\mu_1
\eeq
It can be easily verified that $H_{\pm}$ are shape invariant and for $\lam_1>0$ and $\mu_1>-\a$, supersymmetry is broken. However by a reflection of the parameter $\lam_1$ \cite{gang1} one gets
\beq
V_+(x,\lam_1,\mu_1)=V_-(x,-\lam_1,\mu_1+\a)+4\lam_1\mu_1+2\a\lam_1+2\mu_1+\a
\eeq
where $V_-(x,-\lam_1,\mu)$ admits a zero energy state and is of the same form as the first example of Table 1 with different parameters. Consequently it can be treated in a similar way.

We would like to note that there are several other shape invariant potentials for which  potential algebras can be constructed. Since they can be found out in a similar way, we have omitted the details and have presented the results in Table 1. \\ 

\section{Discussion}
Following the method of ref \cite{balan} it has been shown here that symmetry algebras belonging to the class of potential algebras can be constructed for a class of shape invariant PDMSE's. We feel that potential algebras similar to the ones considered in ref \cite{gan} may also be constructed at least for some shape invariant potentials within the framework of PDMSE. We would also like to point out that here we have not considered the most general translationally shape invariant potentials. For instance let us consider Example 1 and take the superpotential $W(x) = (\lam_1 x+\mu_1)$ and the deforming function $f(x)=(\a x^2+2\b x+1)$. In this case the partner potentials are given by $V_{\pm}(x) = \lam_1(\lam_1\pm\a)x^2+2\lam_1(\mu_1\pm\b)x+\mu_{1}^{2}\pm\lam_1$. These two parameter potentials can be shown to be shape invariant and the parameter relations are given by $\lam_{n+1}=\lam_n+\a$ and $\mu_{n+1}=\f{\lam_n\mu_n+2\b \lam_n+n^2\a\b}{\lam_n+\a}$. Thus the difference between two successive $\lam$'s is a constant while it is not so for two successive $\mu$'s. Similar generalizations can be carried out for some of the potentials listed in Table 1 \cite{quesne1}. We feel It would be interesting to examine whether or not this type of shape invariant potentials as well as the self similar potentials \cite{dutt} can be treated within the present framework.
\begin{center}  
\small
\begin{tabular}{|   l|   l|  l|    l |     l|   l   |  l  | l |   }

\hline $W(x,\boldsymbol{\lambda})$    &$f(x;\boldsymbol{\alpha})$    & $V_{-}=W^2-fW'$  & $\mbox{Parameter~relations}$  & $R(\boldsymbol{\lambda_n})$ &$\mbox{Potential algebras}$ \\

\hline    $\frac{\lambda_1}{x}+\mu_1 x$   &  $\a x^2+1$    & $\frac{\lambda_1(\lambda_1+1)}{x^2}+$  &$\lambda_n=\lambda_{n-1}-1$  & $-4(\a \lambda_n-\mu_n-\a)$ &$$ \\
$\lam_1<0$ &$\a>0$ &$\mu_1(\mu_1-\a)x^2+2\lambda_1 \mu_1$ &$\mu_n=\mu_{n-1}+\a$ &$$  &$[B_{-},B_{+}]=R(\lambda_0)$\\
$$ &$$ &$+\lambda_1 \alpha-\mu_1$ &$$ &$$  &$[B_{\pm},R(\lambda_0)]=\pm 8\a B_{\pm}$\\


\hline $\lambda_1 \tan(x)$ &$1+\alpha \sin^2(x)$ &$\lambda_1(\lambda_1-\alpha-1)sec^2(x)$ &$\lambda_n=\lambda_{n-1}+\alpha+1$ &$2\lambda_n+(\alpha+1)$  &$$\\
$\lam_1>0$ &$\a>-1$ &$-\lambda_1(\lambda_1-\alpha)$ &$$ &$$  &$[B_{-},B_{+}]=R(\lambda_0)$\\
$$ &$$ &$$ &$$ &$$  &$[B_{\pm},R(\lambda_0)]=$\\
$$ &$$ &$$ &$$ &$$  &$\pm 2(1+\alpha)B_{\pm}$ \\

\hline $\lambda_1 \cot(x)$ &$1+\beta \sin^2(x)$ &$\lambda_1(\lambda_1+1) cosec^2(x)$ &$\lambda_n=\lambda_{n-1}-1$ &$-(1+\beta)(2\lambda_n-1)$  &$$\\
        $\lam_1<0$ &$\beta >-1$ &$-\lambda_{1}^{2}+\beta \lambda_1$ &$$ &$$   &$[B_{-},B_{+}]=R(\lambda_0)$\\
        $$ &$$ &$$ &$$ &$$  &$[B_{\pm},R(\lambda_0)]=$\\
        $$ &$$ &$$ &$$ &$$  &$\pm 2(1+\beta)B_{\pm}$\\

\hline $\lambda_1 \coth(x)$ &$1+\alpha \sinh^2(x)$ &$\lambda_1(\lambda_1+1)cosech^2(x)$ &$\lambda_n=\lambda_{n-1}-1$ &$(1-\alpha)(2\lambda_n-1)$ &$$\\
 $\lam_1<0$ &$\alpha >0$ &$+\lambda_1(\lambda_1+\alpha)$ &$$  &$$  &$[B_{-},B_{+}]=R(\lambda_0)$\\
 $$ &$$ &$$ &$$ &$$  &$[B_{\pm},R(\lambda_0)]=$\\
 $$ &$$ &$$ &$$ &$$  &$\pm (\alpha-1)B_{\pm}$\\

\hline 
\end{tabular}
\label{table1}
\small
\end{center}
~~~~\textbf{Table 1:} Potential algebras associated with various shape invariant potentials.


\ed